\documentclass{aa}  

\usepackage{graphicx}
\usepackage{txfonts}
\usepackage{natbib}
\usepackage{amsmath}
\usepackage[dvipsnames]{xcolor}
\bibpunct{(}{)}{;}{a}{}{,} % to follow the A&A style
\begin{document}

   \title{A powerful machine learning technique to extract proton core, beam and $\alpha$-particle parameters from velocity distribution functions in space plasmas}

   %\subtitle{Wave-particle in an ion cyclotron wave}

   \author{D. Vech \inst{1}, M.~L. Stevens\inst{2}, K.~W. Paulson\inst{2}, D. M. Malaspina\inst{1,3},  A.~W. Case\inst{2}, K.~G. Klein\inst{4},
          \and
         J.~C. Kasper\inst{5,6}
          }

   \institute{Laboratory for Atmospheric and Space Physics, University of Colorado, Boulder, CO, USA\\
              \email{daniel.vech@lasp.colorado.edu}
         \and
                 Smithsonian Astrophysical Observatory, Cambridge, MA 02138 USA\\
        \and
                 Astrophysical and Planetary Sciences Department, University of Colorado, Boulder, CO, USA\\
        \and
             Lunar and Planetary Laboratory, University of Arizona, Tucson, AZ 85719, USA\\
             \and
           BWX Technologies, Inc., Washington DC 20002, USA\\
             \and
        Climate and Space Sciences and Engineering, University of Michigan, Ann Arbor, MI 48109, USA\\
             }

   \date{Received DD/MM/YYYY; accepted DD/MM/YYYY}

% \abstract{}{}{}{}{} 
% 5 {} token are mandatory
 
  \abstract
   {The analysis of the thermal part of velocity distribution functions (VDF) is fundamentally important for understanding the kinetic physics that governs the evolution and dynamics of space plasmas. However, calculating the proton core, beam and $\alpha$-particle parameters for large data sets of VDFs is a time consuming and computationally demanding process that always requires supervision by a human expert.}
   {We developed a machine learning tool that can extract proton core, beam and $\alpha$-particle parameters using  images (2-D grid consisting pixel values) of VDFs.}
   {A database of synthetic VDFs is generated, which is used to train a convolutional neural network that infers bulk speed, thermal speed and density for all three particle populations. We generate
   a separate test data set of synthetic VDFs that we use to compare and quantify the predictive power of the neural network and a fitting algorithm.}
   {The neural network achieves significantly smaller root-mean-square errors to infer proton core, beam and $\alpha$-particle parameters than a traditional fitting algorithm.}
   {The developed machine learning tool has the potential to revolutionize the processing of particle measurements since it allows the computation of more accurate particle parameters than previously used fitting procedures.}

   \keywords{solar wind -- waves -- turbulence -- thermal ions
               }

   \maketitle
%
%-------------------------------------------------------------------

\section{Introduction}

The solar wind is a hot, tenuous plasma propagating away from Sun's
surface \citep[e.g.][]{wolfe1966observations}. Determining the properties of the particle populations (protons, electrons and $\alpha$-particles) in the solar wind is fundamentally important for describing the radial evolution of the plasma \citep[e.g.][]{richardson2003radial}, dissipation of turbulent energy \citep[e.g.][]{coleman1968turbulence}, the onset of plasma instabilities \citep[e.g.][]{kasper2002wind} and wave-particle interaction \citep[e.g.][]{howes2017diagnosing}. Protons constitute 95\% of the number density of the solar wind \citep{feldman1978long} and are typically well described by a combination of a peaked function-- most often a Maxwellian velocity distribution function (VDF) for the primary proton population, and a separate VDF for the smaller secondary population that is differentially flowing with respect to the primary population \citep[e.g.][]{tu2004dependence, alterman2018comparison}. The primary proton population is usually referred to as the core and the smaller one the beam. The second most common ion species is $\alpha$-particles that constitute $\approx$4\% of the solar wind number density. Velocity distribution functions in the solar wind are measured in situ by particle detectors such as electrostatic analyzers \citep[ESA, e.g.][]{sauvaud2008impact} and Faraday Cups \citep[e.g.][]{ogilvie1995swe, kasper2016solar}.

Faraday Cups have been used for space plasma exploration for over half a century and they have flown on several spacecraft including NASA's Parker Solar Probe (PSP) mission \citep{fox2016solar, kasper2016solar}. A Faraday Cup measures the currents due to charged particles reaching the collector plates. The discrimination of charged particles and energy per charge determination is based on a time-varying positive potential, which chops a selected portion of the charged particle flux. A capacitor integrates the chopped current from each collector plate in a fixed time interval and the resulting voltage is then converted to a digital signal \citep{ogilvie1995swe}. The voltage resolution of Faraday Cups is typically between 5\% and 13\%.

The key fluid parameters of the VDFs (velocity, density, thermal speed) are derived by fitting the observations to a bi-Maxwellian function. This is an iterative process that aims to minimize the residuals in the fit. The fitting procedure is particularly difficult or potentially impossible when there is a significant overlap between the core, beam and $\alpha$-particle populations, which leaves large uncertainties in the computed fitting parameters. When fitting to multiple populations is impossible, cruder estimates for density, thermal speed, and bulk speed have to be made, such as partial velocity moments of the distribution as measured. The weakness of partial moments is that they do not extrapolate to account for the tails or the overlapping portions of the distributions.

Some Faraday Cup experiments use multiple cups with different orientations such as Voyager \citep{bridge1977plasma}, Wind \citep{ogilvie1995swe}, Europa Clipper \citep{grey2018europa} or a spinning platform such as Wind \citep{ogilvie1995swe} and IMP-8 \citep{lazarus1996solar} for additional geometrical constraints to break the degeneracy between populations. The problem of confusion between multiple populations is especially difficult for experiments that measure a single projection of the solar wind VDF: single-sensor, three-axis stabilized systems like PSP or Deep Space Climate Observatory \citep{burt2012deep}.

Another difficulty is that modern Faraday Cups are capable of high cadence ($\approx$0.05-1 sec) sampling of the solar wind proton VDFs leading to the generation of massive data sets. The Faraday Cup \citep[Solar Probe Cup, SPC, ][]{case2020solar} of PSP obtains over $10^6$ proton VDFs during each solar encounter period. The processing of this huge data set is a time consuming, computationally demanding process that requires supervision by a human expert.

Recently, Machine Learning (ML) techniques have been applied to several problems in space plasma and solar physics such as predicting solar flares \citep[e.g.][]{chen2017convolutional, zheng2019solar, li2020predicting}. One particular ML technique that has played an important role in these advancements is the Convolutional Neural Network (CNN). These networks are most commonly applied to image recognition since they have exceptional ability to identify patterns and features. The output of a CNN can be a discrete variable (a variable that can only take on specific values such as predefined image labels) or a continuous variable, which can take on an unlimited number of values. Given the success of previous studies, we ask the question: is it possible to infer proton core, beam and $\alpha$-particle parameters using images (2-D grid consisting pixel values) of VDFs? Such a predictive tool could have immense applications for plasma investigations by significantly reducing the time and effort needed to post-process the data and obtain accurate particle fits.

In this paper, we develop a powerful ML technique using a Convolutional Neural Network, which can infer proton core, beam and $\alpha$-particle parameters using only images of VDFs as the input. We generate a large data set of  synthetic VDF images where the core, beam and $\alpha$-particle parameters are randomly assigned and the VDFs' characteristics (signal-to-noise ratio, width of the voltage bins) mimic the measurements of SPC. We train the CNN and compare the inferred particle parameters with the ones obtained with the fitting algorithm of the SPC instrument team. Our results suggest that the predictive power of the CNN is significantly better than previous fitting procedures even when the VDFs are affected by substantial noise.

\section{Training data set}

For the generation of the synthetic VDFs we use the SPC response function for a high-Mach number Maxwellian plasma (Stevens et al., in prep). The SPC instrument team fits this response function to the measured solar wind fluxes and in Section 4 we will use this function to extract the core, beam and $\alpha$-particle parameters from the synthetic VDFs. The SPC response function has the following form. %where the three terms correspond the first order solution, 2$^{nd}$ and 4$^{th}$ order polynomials, respectively.

\begin{equation}
\begin{split}
f_i= \frac{1}{2} \phi(V_i) \Delta V_i \times \\
\Bigl(1+\frac{1}{8} (1-u_*(2+w_*^2)+u_*^2) \frac{q \Delta V_i}{2m w^2}^2+\\
\Bigl(\frac{1}{192} (1-(u_*(4+ 6w_*^2 + 12w_*^4 + 15w_*^6))+\\
3u_*^2(2 + 4w_*^2 + 5w_*^4 )-\\
 u_*^3(4 + 6w_*^2) + u_*^4) \frac{q \Delta V_i}{2m w^2}^4 \Bigr) \Bigr) \\
\end{split}
\end{equation}

where

\begin{equation}
\phi(V_i)=\frac{q^2 A n}{mw\sqrt(2 \pi)} exp -\frac{1}{2} \Bigl(\frac{\sqrt(2qV_i/m)-u}{w}\Bigr)^2
\end{equation}

In Equation 1, $\phi(V)$ is the differential flux, which is a function of modulator voltage $V$, $\Delta V$ is the width of the voltage bins (in units of volt), $q$ is the elementary charge, $w$ is the thermal speed, $m$ is the mass of a proton, $u_*=u/v$ and $w_*=w/v$ where $u$ is the bulk speed of the distribution and $v=\sqrt(2qV/m)$. The $i$ subscript corresponds to the i$^{th}$ energy bin. In Equation 2,  $q^2A/m\sqrt{2\pi}=0.8069$ [pA/V], $A$ is the sensitive area of the collector plate and $n$ is the particle density. To simplify our calculations we assume that $q^2A/m\sqrt{2\pi}=1$ in Equation 2.

For the generation of the synthetic VDFs, we use the range of [300; 5000] V for the modulator voltage, which overlaps with the range used by SPC in nominal operation mode. The energy resolution of SPC varies in the range of 5-10\%. We plot the synthetic VDFs on a grid, which has a constant 5\% energy resolution. For more details on the SPC instrument characteristics, see \cite{case2020solar}.

To make the VDFs more realistic, we add random noise to them. The Maxwellian response function with added noise is given by Equation 3,

\begin{equation}
\begin{split}
f_{noisy}= [(f_{total} + (\Gamma \cdot sin(r_1))) ^2 \\
+ (\Gamma \cdot sin(r_2))^2]^{1/2}\\
+ g0.04f_{total}\\
\end{split}
\end{equation}

where $f_{total}$ corresponds to the sum of the core ($f_{core}$), beam ($f_{beam}$) and $\alpha$-particle ($f_{\alpha}$) flux charge densities (obtained with Equation 1), $r_{1,2}$ are random phases drawn from a uniform distribution with [0; 2$\pi$] boundaries, $g$ is also a random variable drawn from a normal distribution with 0 mean. The $g0.04$ term corresponds to the quantization error, errors associated with the modulator waveform, and any additional noise source that is proportionate to the signal. The $\Gamma \cdot sin(r_{1,2})$ terms model background noises on the two separate components of the SPC signal, which are charged current amplitudes measured respectively in-phase and out-of- phase with the high-voltage modulation waveform \citep{case2020solar}. We use $\Gamma =0.5$ pA, which makes for a reasonable comparison to SPC.

The nine parameters of the distributions (bulk speeds $V_{c,b,\alpha}$, thermal speeds $V_{th, c, b, \alpha}$ and density $n_{c, b, \alpha}$ where $c$, $b$ and $\alpha$ subscripts correspond to the core, beam and $\alpha$-particles, respectively) are drawn randomly from a uniform distribution, whose lower and upper boundaries are listed in Table 1. The ranges were chosen such that they are similar to the solar wind properties observed in the inner-heliosphere by PSP \citep{kasper2019alfvenic} and near 1 AU as well \citep{wilson2018statistical, klein2019solar}. The range of differential flows between particle populations were chosen such that full and partial overlap between particle populations is possible, which is a challenging problem for any fitting algorithm.

\begin{table}[h!]
\centering
    \label{tab:table4a5}
   \begin{tabular}{l|r|r|r} % <-- Changed to S here.
      Parameter & Range of possible values \\
      \hline
      $V_{c}$ & [400; 600] km/s \\
      $V_{c, th}$ & [20; 80] km/s\\
      $n_{c}$ & [1; 100] cm$^{-3}$\\
      $V_{b}$ & $V_{c}$+[10; 120] km/s\\
      $V_{b, th}$ & [30; 90] km/s \\
      $n_{b}$ & $n_{c}\cdot$[0.05; 1] cm$^{-3}$\\
      $V_{\alpha}$ & $V_{c}$+[50; 150] km/s\\
      $V_{\alpha, th}$ & [30; 90] km/s \\
      $n_{\alpha}$ & $(n_{c}+n_{b})\cdot$[0.01; 0.05] cm$^{-3}$\\
    \end{tabular}
        \caption{The ranges of possible values for the core, beam and $\alpha$-particle populations.}
\end{table}

In the solar wind the three particle populations are not always simultaneously observed, therefore for the training and testing of the CNN, we generate three groups of images ($10^4$ images in each category) with one, two and three Maxwellian response functions. Figure 1 shows an example of the synthetic VDFs, which were randomly generated and includes three particle populations. The bottom panel shows an actual input image for the training of the CNN where the particle populations have been summed up and noise was added to the resulting distribution.

For generating the image data set, the VDFs are plotted on a fixed scale along the X axis (200 to 1000 km/s), which ensures that the CNN learns to associate higher speeds with VDFs that are positioned closer to the right side of the image and vice versa for slower speeds. The Y axis of the plot is logarithmic and upper and lower thresholds (10$^{-0.3}$-10$^{2.5}$) are chosen such that all VDFs fit in the image box.

\begin{figure}
 %   \figurenum{4}
    \centering\includegraphics[width=1\linewidth]{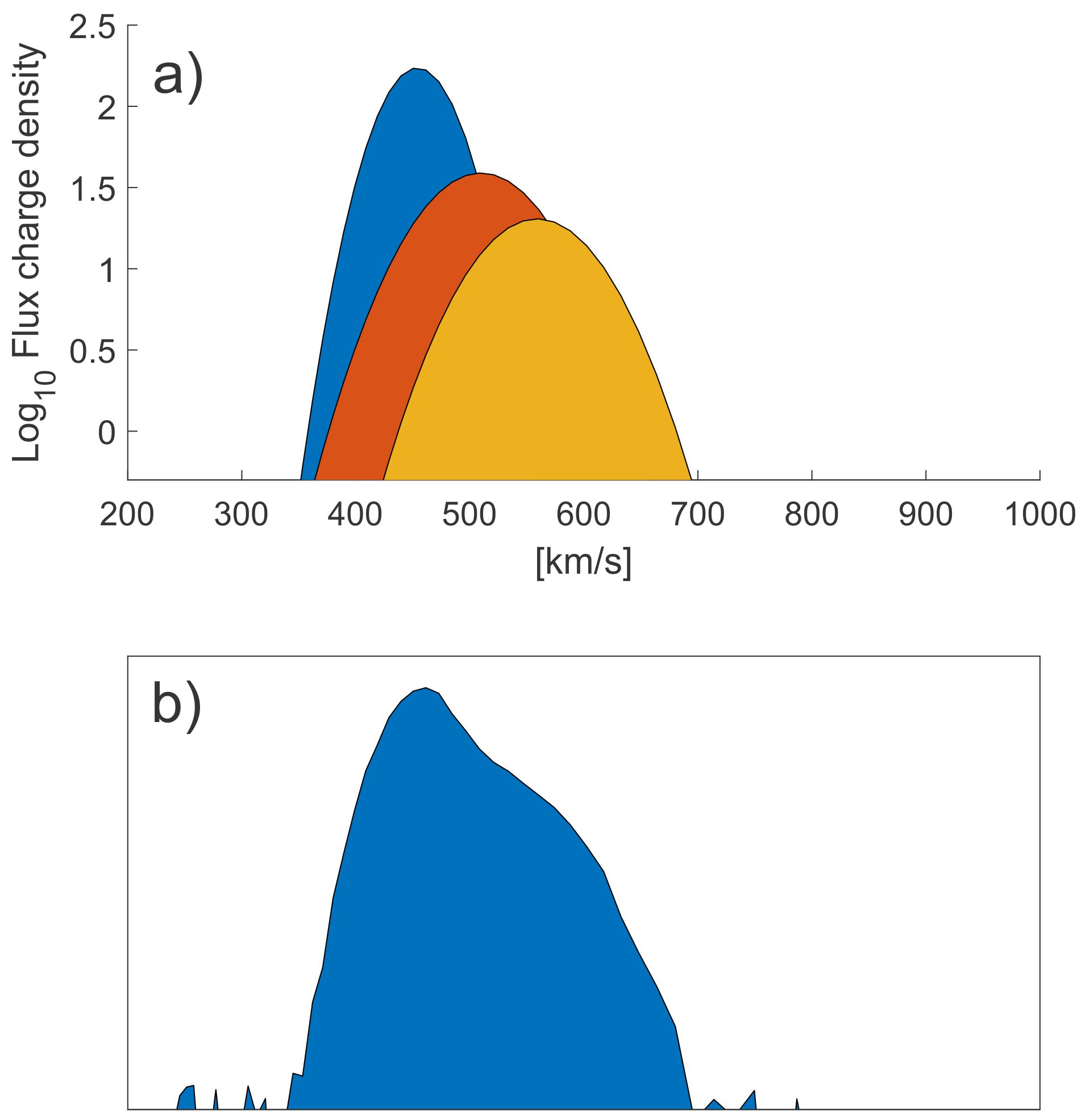}
%\plotone{figure1_landscape.eps}
\caption{Top panel shows one of the 10$^4$ randomly generated core (blue), beam (red) and $\alpha$-particle (yellow) distributions. The bottom panel (actual input image for the CNN) shows the sum of these three particle populations with added noise.}
  \label{fig:PKa2}
\end{figure}

%The Y axis of the plot is automatically adjusted so the peak flux charge density of the VDF reaches 95\% of the height of the image box. We note that due to this plotting method the CNN is not able to predict proton core and beam densities separately. However, we will show in the next section that predicting core-beam density ratio is possible with good accuracy.

%From an image recognition point of view, using a fixed Y axis is not optimal since VDFs with high density would fill up the entire image box while VDFs with low density may make up only a few pixels in the image. Extracting features from them is difficult and the prediction accuracy is low. Using adjustable Y axis means that all VDFs cover approximately the same area of the image box. We note that logarithmic Y axis was also tested, however it led to large inaccuracies in the predictions potentially due to the fact that the pixel size of important features in the VDFs were reduced.

\section{Training and testing the CNN}
Our goal is to develop a data processing pipeline that takes the input image of a VDF and extracts the corresponding particle parameters. This includes two steps (see Figure 2): in the first step, our goal is to train a CNN, which is capable of identifying the correct number of particle populations in a given VDF. In the second step, we will train separate neural networks for extracting bulk speed, thermal speed and density of VDFs with 1, 2 and 3 particle populations.

The CNN used in this paper was obtained from an online repository and is available on (https://www.mathworks.com/help/deeplearning/ug/train-a-convolutional-neural-network-for-regression.html). This network was originally designed to classify images of handwritten numbers.

For step one, we train an image classification CNN on 27000 labelled VDF images (9000 images selected from each category) where the labels are "1", "2" and "3" corresponding to the number of particle populations in a VDF. The remaining 3000 VDFs (1000 from each category) are used for testing the CNN's performance. The used model parameters are summarized in Table 2. For the image classification problem we use Stochastic Gradient Descent (SGD) optimizer \citep{bottou2012stochastic}. The optimizer is an algorithm, which is used to change the parameters of the network such as weights and learning rate to reduce the losses and find the local minimum of a differentiable function. The learning rate defines how big or small the step-sizes are that the optimizer takes into the direction of the local minimum. If the learning rate is too large, there is a risk that the optimizer misses a local minimum. On the other hand, very low learning rate means that the training process takes substantial amount of time. We found that decreasing the initial learning rate did not lead to further improvements in the accuracy of the classification. We also found that increasing the number of epochs (the number of times that the learning algorithm works through the entire training dataset) above 20 did not lead to further improvements in the prediction accuracy. Various image sizes (50x50, 75x75 pixels) were tested, however we found that increasing the image size above 100x100 pixels leads to no further improvement in the model accuracy therefore the generated VDF images are re-scaled to this size.

\begin{table}[h!]
\centering
    \label{tab:table46}
   \begin{tabular}{l|r|r|r} % <-- Changed to S here.
      Model parameter & Value \\
      \hline
      Optimizer & SGD, Adam \\
      Initial Learning Rate & 10$^{-3}$\\
      Learn Rate Drop Factor & 0.1\\
      Epochs & 20\\
    \end{tabular}
        \caption{Hyperparameters of the CNNs. For image classification we used Stochastic Gradient Descent (SGD) optimizer, while for the regression problem the Adam optimizer \citep{kingma2014adam} led to the best results. The other parameters were the same for both type of networks.}
\end{table}

\begin{figure}
 %   \figurenum{4}
    \centering\includegraphics[width=1\linewidth]{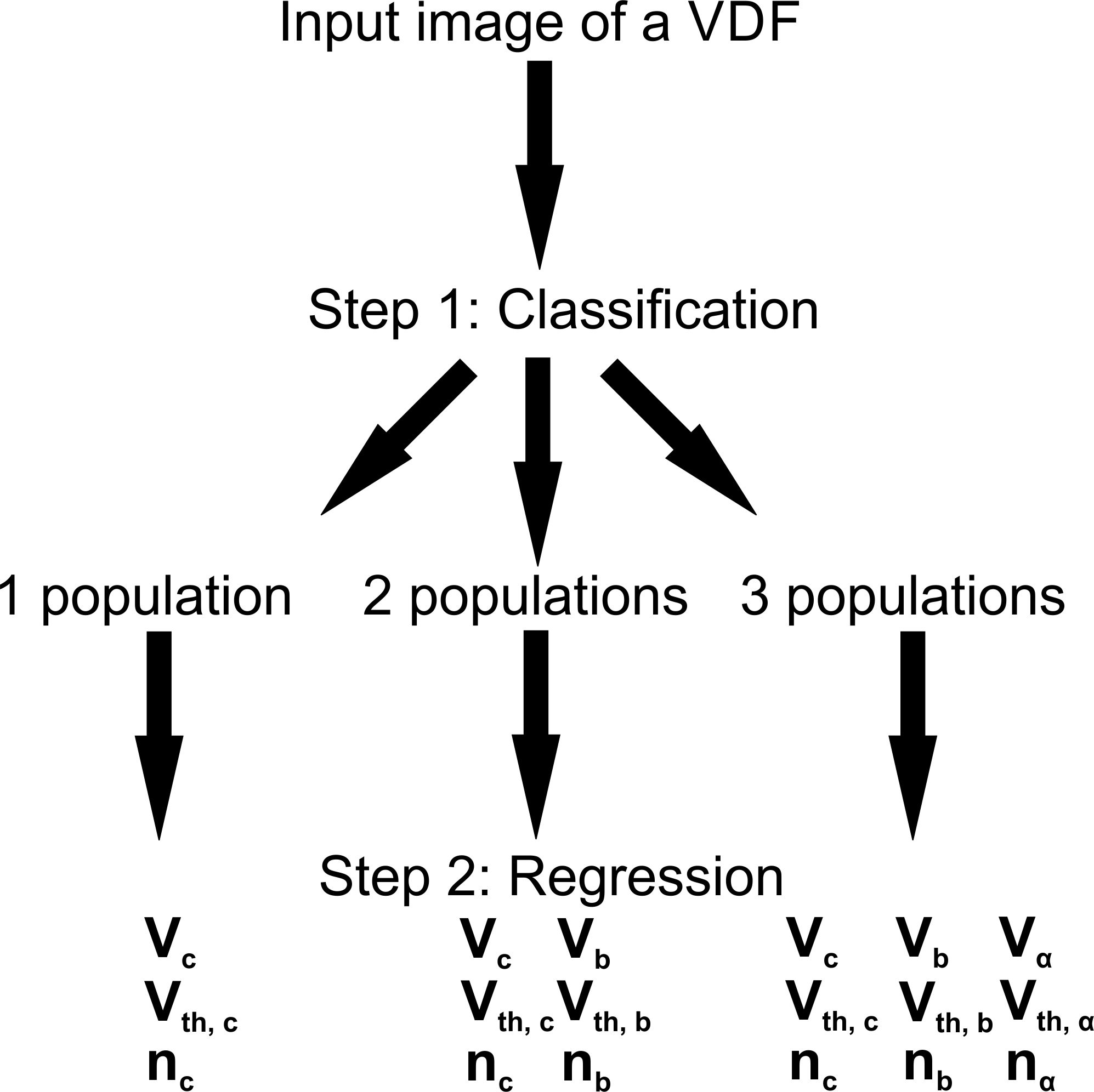}
%\plotone{figure1_landscape.eps}
\caption{Workflow of the proposed approach to process VDFs.}
  \label{fig:PK1}
\end{figure}

\begin{figure}
    \centering\includegraphics[width=1\linewidth]{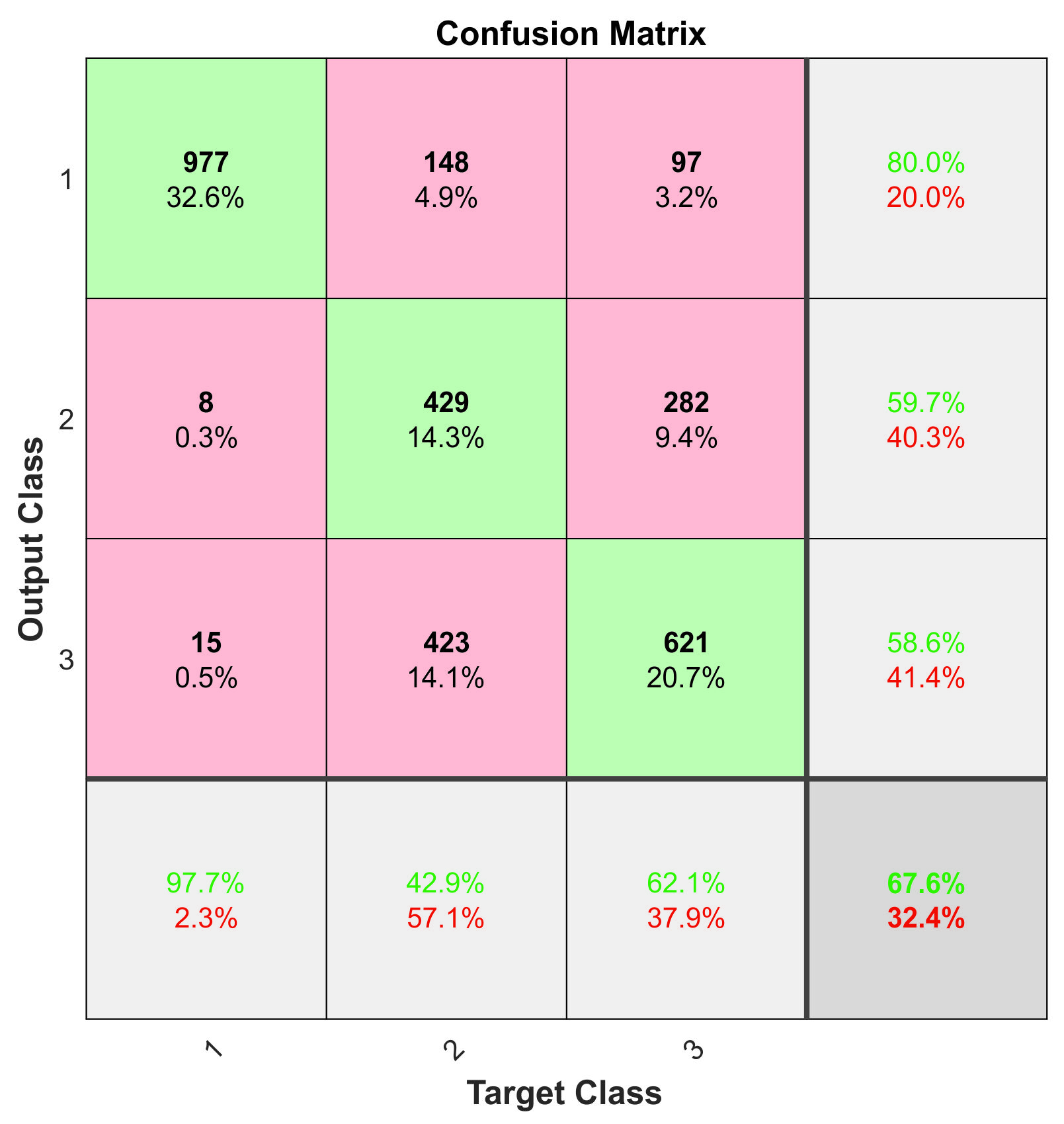}
\caption{Confusion matrix of the classification problem.}
  \label{fig:PK2}
\end{figure}

Figure 3 shows the confusion matrix of the CNN using the 3000 test images. In the confusion matrix the rows correspond to the inferred class (Output Class; "1", "2" and "3" corresponding to the number of particle populations in a VDF) and the columns correspond to the true class (Target Class). The correctly classified observations are in the diagonal of the matrix. Both the right column and bottom row are important to asses the accuracy of the classification. The bottom row shows the percentages of all the examples belonging to each class that are correctly (true positive, green) and incorrectly (false negative, red) classified. The right column shows the percentage of cases when the CNN predicted 1 (80.0\% correct), 2 (59.7\% correct) and 3 (58.6\% correct) particle populations and the prediction was correct. The values in the bottom row and right column do not necessarily have to be the same. For example, if our network inferred 1 particle population for all the 3000 VDFs, then there would be 100\% in the bottom left corner of the confusion matrix and 33.3\% in the top right corner, respectively, which would be very poor classification.

The overall accuracy of the CNN is 67.6\%. The network has excellent ability to distinguish VDFs with 1 particle population from the other two categories (97.7\% true positive rate). The most challenging problem is distinguishing between 2 and 3 particle populations, in particular when the Target Class is 2 (42.9\% true positive rate).

In the second step, we train separate CNNs to extract particle parameters from VDFs with one, two and three particle populations. For this regression problem, we use a CNN, which has only one output variable; this significantly reduces the training time and complexity of the model. This means that we train 3, 6 and 9 separate CNNs for VDFs with 1, 2 and 3 particle populations, respectively. We have tested CNNs with three output parameters (e.g. bulk speed, thermal speed, density of a particle population), however, we found slight ($\approx$20\%) increase in the root-mean-square error (RMSE) of the inferred values. This might be explained by the fact that the speed, thermal speed and density are randomly generated (Table 1) therefore the CNN could not identify correlations between them. It is also possible that the RMSE increased due to the fact that we increased the number of free variables in the CNN.

Before the training process, the output parameter is normalized to its z-scores ($Z=(y-\mu)/\sigma$ where $y$ is the output parameter and $\mu$ and $\sigma$ are the mean and standard deviations of $y$, respectively), which is a standard ML approach to increase the numerical stability of the model \citep[e.g.][]{NAWI201332}. The output of the regression network is a dimensionless quantity ($Z$), which is transformed back to physical units with $\sigma$ and $\mu$ (i.e. $y=Z\sigma+\mu$), which are obtained from the training data. We split the initially generated $10^4$ images from each of the three categories into a training (9000 images) and testing (1000 images) sets, respectively and train separate CNNs for all the 18 parameters (3+6+9) in total. During the training process the goal of the CNN is adjusting its model parameters so it can achieve the lowest RMSE to infer the single output parameter. We found that for all parameters the CNN converged to a steady RMSE value after 10-12 epochs; between epochs 12-20 the RMSE error fluctuated by $\pm$2-3\% with respect to the previously achieved RMSE value and no further improvement was achieved.

After training both the classification and regression networks, our data processing pipeline shown in Figure 2 can be tested. We take the 3000 test VDFs (1000 from each of the three categories; these VDFs were not used for training any of the networks), use the previously trained CNN to classify them into 3 categories. Then, we use the separate CNNs to infer the bulk speed, thermal speed and density of each particle species. We note that we compare the inferred parameters with the ground truth parameter (e.g. parameter used for generating the VDF) even if the number of particle populations is incorrectly inferred. For example, if a VDF includes 3 particle populations but the classification networks infers that only 2 particle populations are present, then our comparison will be based on the inferred two particle populations.

The result of this test is summarized in Table 3 where we listed the RMSE and Pearson correlation coefficients between the inferred and ground truth values, respectively. In the CNN the quality of the inferred values is not evaluated based on a test such as the chi-square statistics, which can give misleading results in the case of the  traditional fitting algorithm (e.g. low chi-square value despite large deviations of the fit from the data.).

The training dataset was based on "reasonable" VDFs, which were generated with noise amplitudes typically observed in the SPC data. To test the response of our technique to "unreasonable" VDFs, we generated 3000 VDFs (with the parameters listed in Table 1) where we increased the amplitude of the background noise from 0.5 (see Equation 3) to 20. These VDFs are indistinguishable from pure noise and does not show any resemblance to the example in Figure 1. We found that on this data set the CNN inference had no correlation (Pearson correlation coefficients of $\approx$0) with the ground truth values, however, all the 9 inferred parameters were in the range of the training data set (Table 1). This test shows that the CNN has to be carefully used on real data since in the case of extreme noise the inferred values are still "reasonable" and do not stand out from the rest of the data as being erroneous.

%\begin{table}[h!]
%\centering
%    \label{tab:table45}
%   \begin{tabular}{l|r|r|r} % <-- Changed to S here.
%      Parameter & RMSE & Median \% error & Corr. coef. \\
%      \hline
%      $V_{c}$ & 26.3 km/s & 2.8\% & 0.91\\
%      $V_{b}$ & 36.3 km/s & 8.8\% & 0.86\\
%      $V_{\alpha}$ & 33.2 km/s & 14.4 \% & 0.85\\
%      $V_{c, th}$ & 7.7 km/s & 3.9\% & 0.88\\
%      $V_{b, th}$ & 12.4 km/s  & 13\% & 0.66\\
%      $V_{\alpha, th}$ & 16.4 km/s  & 28.4\% & 0.32\\
%      $n_{c}$ & 10.8 cm$^{-3}$ & 4.3\% & 0.92\\
%      $n_{b}$ & 10.5 cm$^{-3}$  & 17.6\% & 0.90\\
%      $n_{\alpha}$ & 1.7 cm$^{-3}$  & 17.23\% & 0.90\\
%    \end{tabular}
%        \caption{RMSE, median percentage error and correlation %coefficient of the predicted particle parameters by the CNN.}
%\end{table}

%\begin{table}[h!]
%\centering
%    \label{tab:table45}
%   \begin{tabular}{l|r|r|r} % <-- Changed to S here.
%      Parameter & RMSE & Corr. coef. \\
%      \hline
%      $V_{c}$ & 26.6 km/s & 0.91\\
%      $V_{b}$ & 33.1 km/s &  0.88\\
%      $V_{\alpha}$ & 33.8 km/s &  0.85\\
%      $V_{c, th}$ & 7.35 km/s &  0.90\\
%      $V_{b, th}$ & 10.8 km/s  &  0.74\\
%      $V_{\alpha, th}$ & 17.1 km/s  &  0.25\\
%      $n_{c}$ & 10.3 cm$^{-3}$ &  0.93\\
%      $n_{b}$ & 11.2 cm$^{-3}$  &  0.89\\
%      $n_{\alpha}$ & 1.06 cm$^{-3}$  &  0.81\\
%    \end{tabular}
%        \caption{RMSE and correlation coefficient of the predicted %particle parameters by the CNN.}
%\end{table}

\begin{table}[h!]
\centering
    \label{tab:table4543}
   \begin{tabular}{l|r|r|r|r|r} % <-- Changed to S here.
        & RMSE (CNN) & RMSE (Fit) & CC (CNN) & CC (Fit)\\
      \hline
      $V_{c}$ & 26.6 km/s & 28.6 km/s & 0.91 & 0.90\\
      $V_{b}$ & 33.1 km/a & 77.2 km/s & 0.88 & 0.59\\
      $V_{\alpha}$ & 33.8 km/s & 91.0 km/s &  0.86 & 0.85\\
      $V_{c, th}$ & 7.35 km/s & 11.6 km/s &  0.90 & 0.70\\
      $V_{b, th}$ & 10.8 km/s & 36.1 km/s  &  0.74 & 0.26\\
      $V_{\alpha, th}$ & 17.1 km/s &  36.4 km/s  &  0.25 & 0.01\\
      $n_{c}$ & 10.3 cm$^{-3}$ & 23.4 cm$^{-3}$ &  0.93 & 0.86\\
      $n_{b}$ & 11.2 cm$^{-3}$ & 26.0 cm$^{-3}$  &  0.89 & 0.41\\
      $n_{\alpha}$ & 1.06 cm$^{-3}$ & 2.3 cm$^{-3}$  &  0.81 & 0.25\\
    \end{tabular}
        \caption{RMSE and Pearson correlation coefficients (CC) of the inferred particle parameters by the CNN and the fitting algorithm on the 3000 test VDFs.}
\end{table}

%\begin{figure}
 %   \figurenum{4}
%    \centering\includegraphics[width=1\linewidth]{Figure_02.jpg}
%\plotone{figure1_landscape.eps}
%\caption{Predicted and ground truth values for the core and beam bulk speed and thermal speed.}
%  \label{fig:PK}
%\end{figure}

%We note that in our synthetic data set a secondary population is always included therefore the CNN assumes the presence of a beam. In the solar wind, proton beams are not always observed, therefore if the predicted parameters are $V_b\approx V_c$ and $V_{c, th}>V_{b, th}$ there may not be a significant beam in the VDF.

%\textcolor{red}{The prediction of $n_b/n_c$ still needs to be finished} The accurate measurement of the beam density is only possible when there is a small overlap between the core and the beam. For this purpose we selected those VDFs where the ground truth values satisfy the $(V_b-V_c)/V_{c, th}>1.2$ relation. This led to identification of 4790 images that we split into training (90\%) and test (10\%) data set and the CNN was trained to predict the $n_b/n_c$ ratio. The results are shown in Figure 3 and the RMS error is listed in Table 3.

%CNN's ability to extrapolate to a domain. We have tested VDFs that lie outside the domain of the training data set. For this purpose we tested the following ranges
%In order to test the limits of the CNN, we generated a new data set of 3000 VDFs, which are contaminated by substantial noise: the  in Equation is set to 0.4 meaning that the magnitude of the Gaussian noise increased by on order of magnitude.

\section{Comparison to traditional fitting algorithm}

The fitted parameters are obtained by nonlinear least-squares regression of the synthetic VDFs to the analytic model of the SPC instrument response (Equation 1) to one, two or three isotropic Maxwellian distributions of inflowing ions. The RMSE and Pearson correlation coefficients of the fitting algorithm on the test data are listed in Table 3. It can be seen that the CNN achieves lower RMSE values and higher Pearson correlation coefficients for all 9 parameters.

In Figure 4 we compare the inferred and the fitted values for the 3000 test VDFs. The top row shows the core parameters: the CNN achieves a minor improvement compared to the fitting algorithm (factor of 1.075 decrease in RMSE) in its computation of bulk speed, while the CNN RMSEs are significantly smaller (factors of 1.57 and 2.27) than the fitting algorithm for the core thermal speed and density. The beam parameters in the second row show that the CNN RMSE values are smaller than the fitting algorithm RMSE values by factors of 2.33, 3.34 and 2.32 for the beam speed, thermal speed and density, respectively. Finally, the bottom row shows the $\alpha$-particle comparison. The $\alpha$-particle speed in the SPC algorithm is determined by shifting the bulk proton speed by the $\alpha$-to-proton mass factor hence the large discrepancies between the SPC values and the ground truth values. The CNN predictions of $\alpha$-particle speed have a similar RMSE value to the proton core and beam speeds. The CNN predictions of $\alpha$-particle thermal speed have the lowest correlation coefficient (0.25, see Table 3) among all the parameters. The CNN $\alpha$-particle density predictions are in good agreement with the ground truth values and the RMSE value is only 1.06 $\#$/cm$^3$.

In Figure 5 we compare the inferred total (core + beam) proton bulk speed, thermal speed and density from the CNN and the fitting algorithm. The RMSE and correlation coefficients are listed in Table 4. By comparing Figure 4 and 5 it can be seen that the CNN does not make significant systematic errors in the effort to separate the core and the beam into two pieces. Additionally, Figure 5 also clarifies that the overshoot from the fitting algorithm for the core density and the core temperature are almost always accompanied by an undershoot in the beam density and temperature.

\begin{table}[h!]
\centering
    \label{tab:table4589}
   \begin{tabular}{l|r|r|r|r|r} % <-- Changed to S here.
        & RMSE (CNN) & RMSE (Fit) & CC (CNN) & CC (Fit)\\
      \hline
      $V_{tot}$ & 26.3 km/s & 26.8 km/s & 0.85 & 0.92\\
      $V_{tot, th}$ & 10.3 km/s & 15 km/s &  0.72 & 0.67\\
      $n_{tot}$ & 7.32 cm$^{-3}$ & 16.3 cm$^{-3}$ &  0.98 & 0.94\\
    \end{tabular}
        \caption{RMSE and correlation coefficients (CC) of the inferred total (core + beam) proton parameters by the CNN and the fitting algorithm on the 3000 test VDFs.}
\end{table}

There are fundamental differences between the two methods that must be considered for the correct interpretation of Figure 4 and 5. The fitting algorithm tested here was designed to process a very broad range of VDFs that occur throughout the orbit of PSP and is able to handle SPC specific noise anomalies while imposing as few assumptions as possible. In contrast, the CNN had the advantage that it was tested on a domain that has the same properties as the training data. Therefore the achieved RMSE values correspond to the best case scenario for the network. Another important difference is that the analytical fitting has the advantage that error bars can be obtained therefore the quality of the fits can be investigated.

The SPC data pipeline imposes constraints of co-movement ($V_c=V_{\alpha}$) and temperature equilibrium ($V_{th, c}=V_{th, \alpha}$) to minimize the $\alpha$-proton confusion in a one-directional way: protons being confused for $\alpha$-particles is acceptable but $\alpha$-particles being confused for protons is not. Similar constraints are not used in the data pipelines of other Faraday Cups such as Wind, which is a major advantage of those data sets. Additionally, a Wind Faraday Cup spectra comprises about 1240 measurements of charged particle flux distributed over 40 different projection geometries, as compared to an SPC spectrum which is about 30 measurements on one geometrical projection. The proposed CNN operates as a stand-alone data processing pipeline, however, it also has the potential to improve the SPC fitting algorithm by inferring the number of particle populations in a given VDF
therefore the constraints on $\alpha$-particles are not required.

%\begin{table}[h!]
%\centering
%    \label{tab:table45}
%   \begin{tabular}{l|r|r|r} % <-- Changed to S here.
%      Parameter & RMSE & Median \% error & Corr. coef. \\
%      \hline
%      $V_{c}$ & 28.6 km/s & 1.19\% & 0.89\\
%      $V_{b}$ & 77.4 km/s & 11.23\% & 0.57\\
%      $V_{\alpha}$ & 88.9 km/s & 29.77 \% & 0.83\\
%      $V_{c, th}$ & 9.4 km/s & 11.32\% & 0.79\\
%      $V_{b, th}$ & 35.4 km/s  & 47.18\% & 0.05\\
%      $V_{\alpha, th}$ & 37.07 km/s  & 72.98\% & -0.02\\
%      $n_{c}$ & 24.18 cm$^{-3}$ & 14.52\% & 0.85\\
%      $n_{b}$ & 23.9 cm$^{-3}$  & 56.41\% & 0.02\\
%      $n_{\alpha}$ & 5.9 cm$^{-3}$  & 96.73\% & 0.2\\
%    \end{tabular}
%        \caption{RMSE, median percentage error and correlation %coefficient of the predicted particle parameters by the fitting %algorithm.}
%\end{table}

%\begin{table}[h!]
%\centering
%    \label{tab:table45}
%   \begin{tabular}{l|r|r|r|r|r} % <-- Changed to S here.
%        & RMSE (Fit) & RMSE (CNN) & CC (Fit) & CC (CNN)\\
%      \hline
%      $V_{c}$ & 28.6 km/s & 0.91\\
%      $V_{b}$ & 77.2 km/s & 0.59\\
%      $V_{\alpha}$ & 91.0 km/s &  0.86\\
%      $V_{c, th}$ & 11.6 km/s &  0.85\\
%      $V_{b, th}$ & 36.1 km/s  &  0.01\\
%      $V_{\alpha, th}$ & 36.4 km/s  &  -0.02\\
%      $n_{c}$ & 23.4 cm$^{-3}$ &  0.86\\
%      $n_{b}$ & 26.0 cm$^{-3}$  &  0.09\\
%      $n_{\alpha}$ & 2.3 cm$^{-3}$  &  0.2\\
%    \end{tabular}
%        \caption{RMSE and correlation coefficient of the predicted %particle parameters by the fitting algorithm.}
%\end{table}

\begin{figure*}
   % \figurenum{4}
   \centering
    \includegraphics[width=1\linewidth]{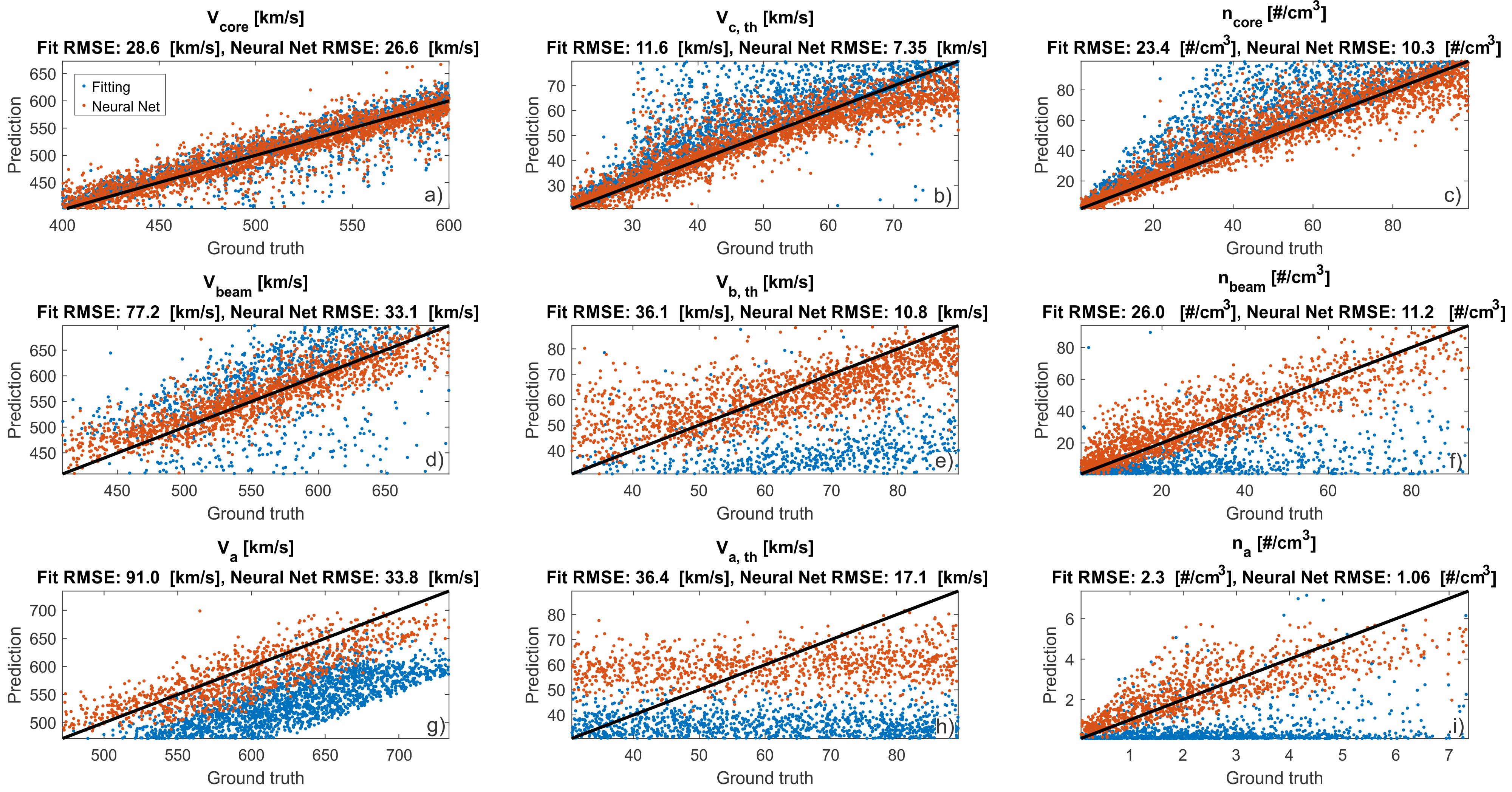}
%\plotone{Figure_04.jpg}
\caption{Inferred and ground truth values for the core, beam and $\alpha$-particle parameters.}
  \label{fig:PK4}
  \end{figure*}

  \begin{figure}
    \centering\includegraphics[width=1\linewidth]{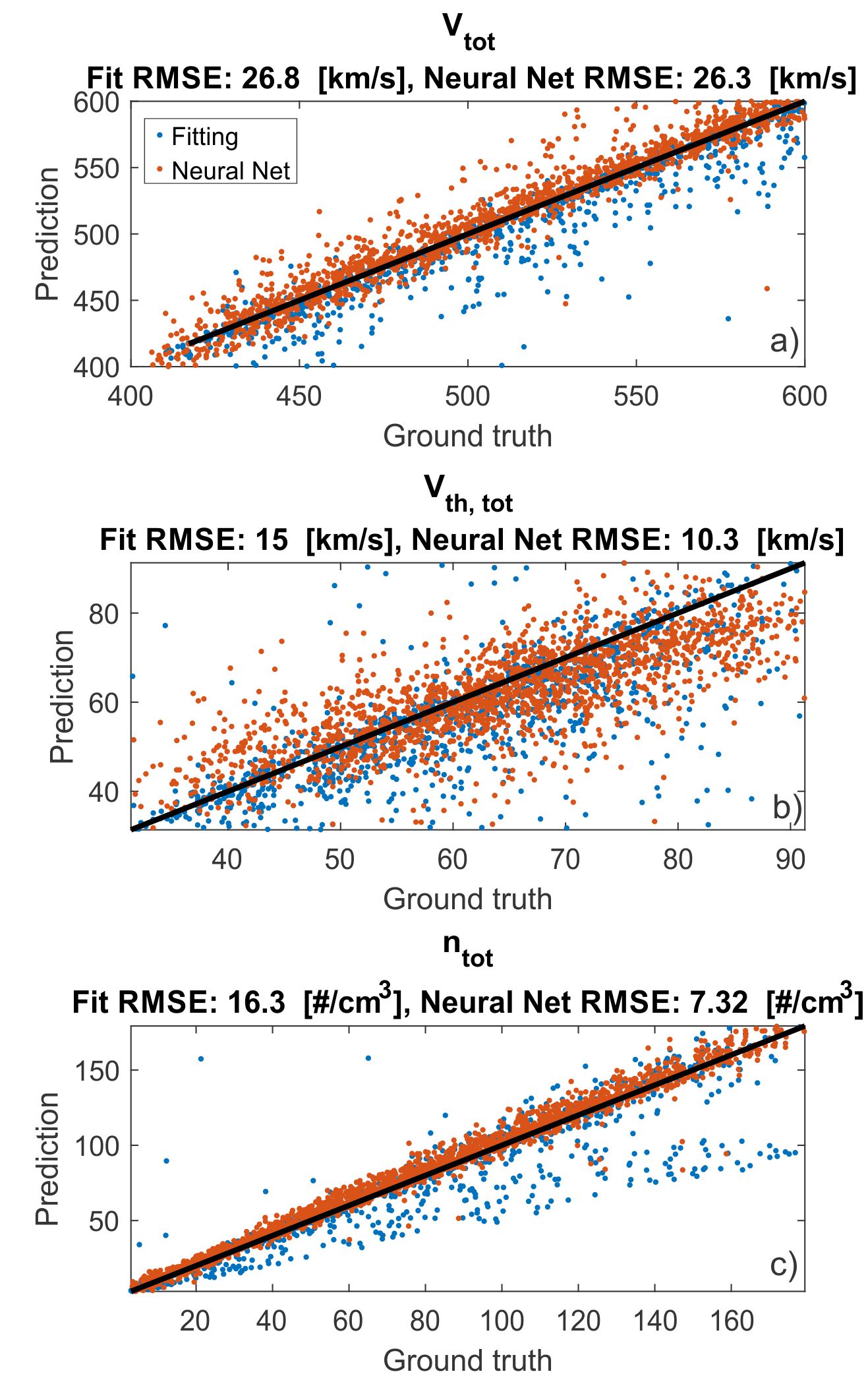}
\caption{Inferred and ground truth values for the total (core + beam) proton population.}
  \label{fig:PK5}
\end{figure}
%  \begin{figure*}[ht!]
%\figurenum{2} \plotone{Figure_}
%\plotone{figure1_landscape.eps}
%\caption{Spectral index of the dissipation range binned in the ($\fb/f_{\rho_s}$, $\beta_e$) plane. The black square marks the region with the steepest spectral indices in the range of $0.1\lesssim\beta_e\lesssim 1$ and $0.12 \lesssim \fb/f_{\rho_s} \lesssim 0.63$.
%}
%  \label{fig:2}
%\end{figure*}

\section{Test with real data}
We have selected 7500 VDFs (19 August 2019 00:00 to 20 August 2019 00:00, orbit \#3) from SPC to test our approach. Only those VDFs were selected that had a data quality flag of 0 ('good data') as determined by the instrument team.

%For generating the image files from the real VDFs, the range of the Y axis has to be carefully chosen so all VDFs fit into the image box (i.e. peak phase space density is less than the upper limit of the Y axis) and all VDFs fill up at least 10\% of the area of the image box (therefore the network has sufficient image pixels to work with). We set the lower and upper limits of the Y axis to 1 and 10$^{1.5}$, respectively. This range is different from the training images (where the range was 10$^{-3}$-10$^{2.5}$) therefore this causes an offset between predicted and actual density and thermal speed values. The predicted density (both beam and core) were corrected with the factors of 0.72$\cdot V_{th, predicted}$ + 11.05 and ($n_{predicted}$+3.90) / 0.40, which were determined based on linear regression between the SPC fits of core thermal speed, density and the CNN predictions.

Figure 6 shows the comparison of the core and secondary population parameters computed with the fitting algorithm and the CNN. In panel g) the ion VDFs are shown; the peak of each VDF was normalized to 1 and plotted on a logarithmic scale between 0.01-1. The largest discrepancy between the two methods is at approximately 00:20-01:20. In Figure 6d-e-f it can be seen that this is the only interval where the CNN predicted 2 particle populations. By inspecting VDFs from these intervals (Figure 7a) we can see that there is a substantial secondary population between 00:20-01:20 and similar features are not present in VDFs outside this interval (Figure 7b), although the fitting algorithm suggested that there is a beam.

\begin{figure*}
   % \figurenum{4}
   \centering
    \includegraphics[width=1\linewidth]{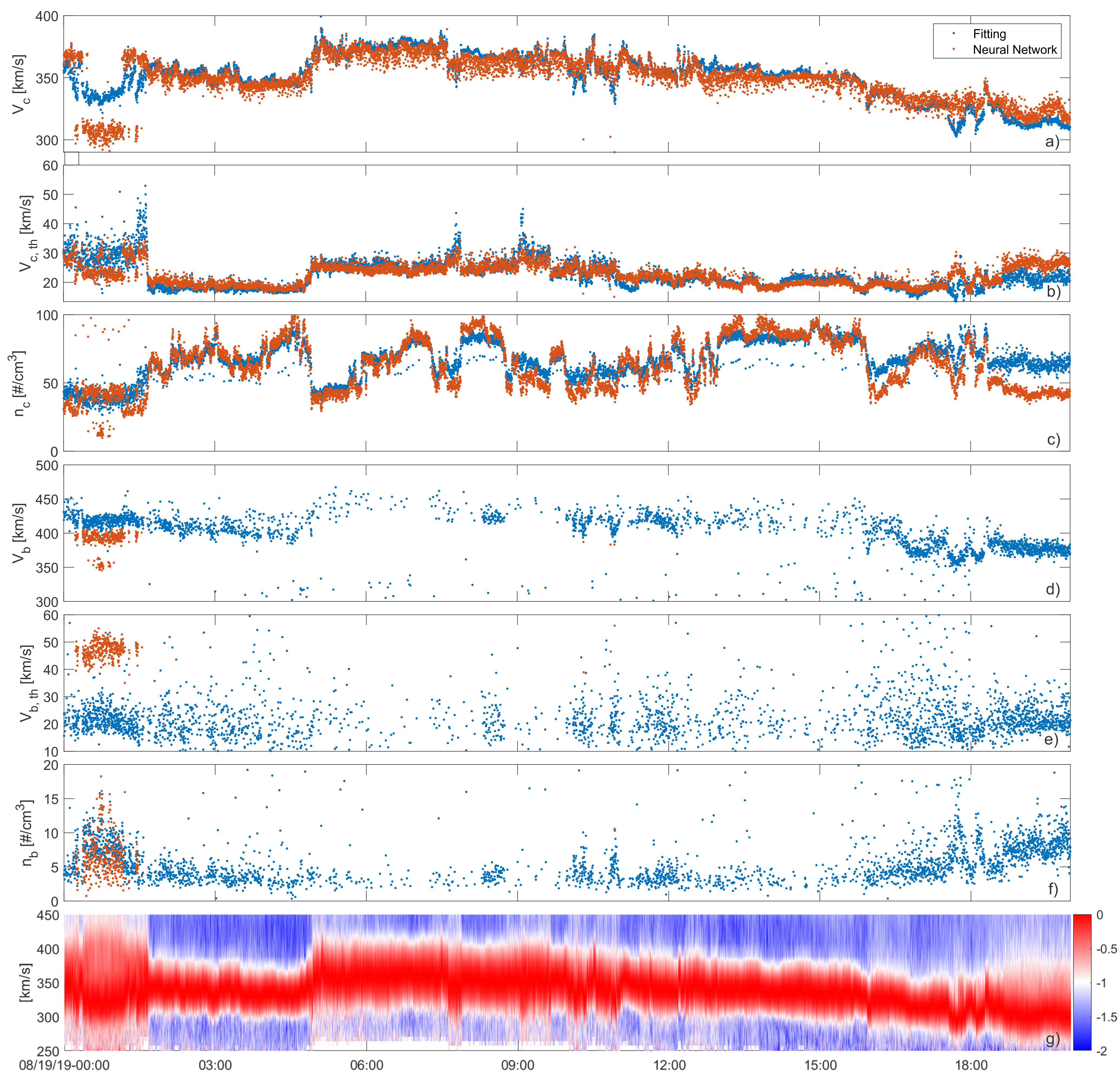}
%\plotone{Figure_04.jpg}
\caption{Comparison of the SPC fitting algorithm and the CNN on a real test data. In pane g) the peak of each VDF was normalized to 1 and the normalized phase space densities are plotted on a logarithmic scale.}
  \label{fig:PK8}
  \end{figure*}

This test shows that the CNN has significantly higher thresholds for labelling a VDF as having two particle populations than the fitting algorithm. In Figure 3 we showed that the CNN has excellent ability (97.7\% true positive rate) to distinguish between 1 vs. 2 (and 3) particle populations, therefore we suggest that outside the 00:20-01:20 interval there is no substantial secondary particle population. The CNN predicted 3 particle populations in only 2 instances out of the 7500 VDFs therefore those cases are not shown.

We have quantified the differences between the fitting algorithm and the CNN in terms of Pearson correlation coefficients and root-mean-square difference, which are shown in Table 5. These results show that the least-square fits offer a reliable way to measure the core protons with SPC data.

\begin{table}[h!]
\centering
    \label{tab:table4500}
   \begin{tabular}{l|r|r|r} % <-- Changed to S here.
        & Root-mean-square difference & Corr. coef. \\
      \hline
      $V_{c}$ & 9.9 km/s & 0.85\\
      $V_{b}$ & 29.6 km/s & 0.16\\
      $V_{c, th}$ & 3.09 km/s & 0.78\\
      $V_{b, th}$ & 25.5 km/s  & 0.35\\
      $n_{c}$ & 10.8 cm$^{-3}$ & 0.87\\
      $n_{b}$ & 3.7 cm$^{-3}$  & 0.13\\
    \end{tabular}
        \caption{Root-mean-square difference and Pearson correlation coefficients for the core and secondary particle population derived with the CNN and the fitting algorithm on the real test data.}

\end{table}

\begin{figure}
 %   \figurenum{4}
    \centering\includegraphics[width=1\linewidth]{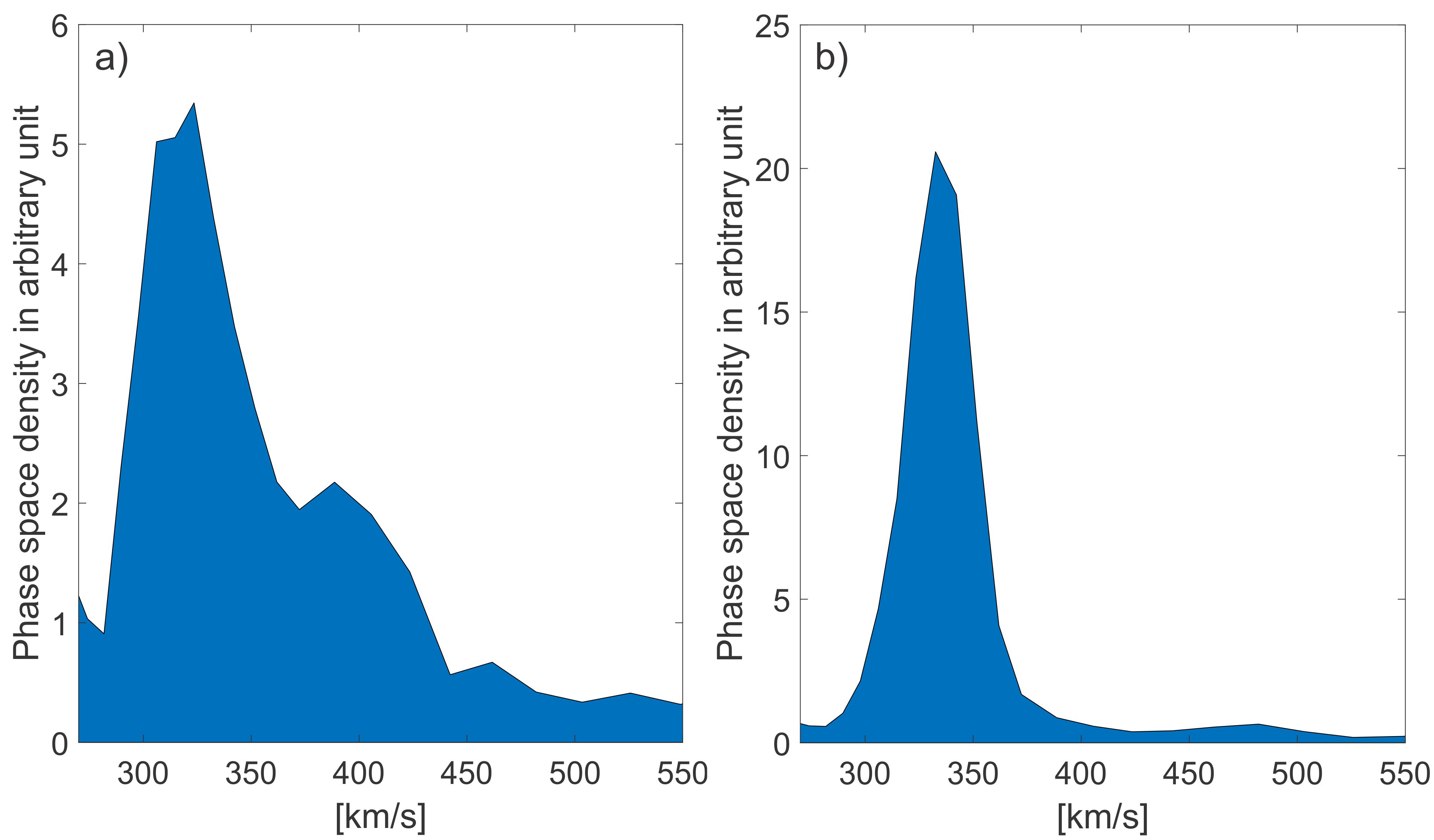}
%\plotone{figure1_landscape.eps}
\caption{a) VDF from 00:44:51, which was identified as having two particle populations by both methods. b) VDF from 03:00:20, where the fitting algorithm suggested both core and a secondary population while the CNN's prediction was that there is only core.}
  \label{fig:PK9990}
\end{figure}

\section{Conclusions}
In this paper, we have developed a new, powerful ML tool to compute parameters of one, two and three Maxwellian response functions. We have demonstrated that the proposed Convolutional Neural Network achieves significantly lower RMSE (Figure 4) than the previously used fitting technique to derive bulk speed, thermal speed and density of particle populations.

The fitting algorithm tested here was designed to work in very diverse conditions with minimal assumptions. In contrast, the CNN is significantly less robust since it was trained on a specific range of solar wind parameters and it works best on a dataset that lies in the domain of the training data set. Additional disadvantage of the CNN is that for an input VDF with extreme noise the network infers "reasonable" (however, incorrect) particle parameters. We suggest that the fitting algorithm should be used in tandem with the CNN to achieve the best performance. This approach would allow to filter out VDFs that are affected by substantial noise by comparing the CNN's inferred core parameters with the fitting algorithm, which are expected to be in good agreement when the noise levels are reasonable. When the two core datasets are consistent, the second and third population can be extracted with the CNN.

More accurate computation of solar wind proton and $\alpha$-particle parameters has major implications for the improved characterization of space plasmas. For example, linear stability analysis is a frequently used technique to understand wave modes and energy exchange between electromagnetic fields and particles in space plasmas. However, large inaccuracies in the input parameters (such as proton thermal speed, differential flow between ion species) may lead to false results suggesting that the plasma is stable (or unstable) against certain instabilities. Our technique will provide more accurate input parameters for linear stability calculations than previous fitting techniques, which will lead to a better understanding of the turbulent cascade including the scale-to-scale energy transfer and the growth rate of unstable wave modes.

Throughout this paper, we used the characteristics of the SPC instrument to demonstrate the feasibility of our approach, however, the technique can be easily adopted to any currently operated or former particle detectors and can be applied to both ions and electrons whose distributions can be modelled through theoretical calculations. %\textcolor{red}{For other space missions such as Wind, our approach will make it possible to reprocess the available large data sets, however, it is difficult predict how much an updated data set could differ from the already available one and speculating about it is far beyond the scope of this study.}

%The most challenging problem for the neural network is to make a distinction between VDFs with two and three particle populations. Incorporating the $\alpha$-particle number density in the calculations could help to clarify if a third particle population is present (e.g. one would only accept the CNN's prediction of three particle populations if the $\alpha$-particle number density exceeds a certain threshold). On PSP the time of flight section of the Solar Probe Analyzers (SPAN-A) enables to sort particles by their mass/charge ratio, permitting differentiation of ion species \citep{kasper2016solar}.
In the training and test VDFs we did not explicitly include a combination of core proton and $\alpha$-particle populations. A sub-set of the core proton and beam VDFs show similarities (in terms of density ratio, temperature ratio, range of differential flow) to the expected core and $\alpha$-particle signatures. Our priority was to use only 3 classes of VDFs in an effort to maximize the accuracy of the classification network and keep the number of free parameters in the CNN low. On PSP the time of flight section of the Solar Probe Analyzers (SPAN-A) enables to sort particles by their mass/charge ratio, permitting differentiation of ion species, which allows the correct determination of the particle species of the second and third particle populations \citep{kasper2016solar}.

We recognize that the CNN technique described in this paper is limited by the fact that it is inherently one-dimensional (i.e., it is using measurements of only the radial dimension of the VDF) and the CNN was trained on typical range of solar wind parameters.  In reality, space plasmas are typically anisotropic with respect to the local magnetic field \citep[e.g][]{hellinger2006solar}, and the flow velocity is not purely radial \citep[e.g][]{kasper2019alfvenic}. Further work will remove these limitations by extending the CNN technique to incorporate solar wind flow angle measurements and three-dimensional VDFs.

\begin{acknowledgements}
KGK was supported by NASA Grant 80NSSC19K0912. D. V. was supported by NASA contract 80NSSC21K0454. D.M. was supported by NASA contract 80NSSC19K0305. The authors thank the Parker Solar Probe team for their support. All data used in this paper is available on the SWEAP data archive: http://sweap.cfa.harvard.edu/pub/data/sci/sweap
\end{acknowledgements}

% for the bibliography, at the end
\bibliographystyle{aa} % style aa.bst

\end{document}